\documentclass[manuscript]{acmart}
\AtBeginDocument{%
  }

\begin{document}

\title{EDU-MATRIX: A Society-Centric Generative Cognitive Digital Twin Architecture for Secondary Education}

\author{Wenjing Zhai}
\email{wendysnake55@163.com}
\orcid{0009-0004-2434-1368}
\affiliation{%
  \institution{The High School Affiliated to Beijing Normal University}
  \city{Beijing}
  \country{China}
}

\author{Jianbin Zhang}
\affiliation{%
  \institution{The High School Affiliated to Beijing Normal University}
  \city{Beijing}
  \country{China}}
\email{zhangjianbin2026@163.com}

\author{Tao Liu}
\affiliation{%
  \institution{Department of Electronic and Communication Engineering, North China Electric Power University}
  \city{Hebei}
  \country{China}}
\email{taoliu@ncepu.edu.cn}

\renewcommand{\shortauthors}{Wenjing Zhai et al.}

\begin{abstract}
Existing multi-agent simulations often suffer from the "Agent-Centric Paradox": rules are hard-coded into individual agents, making complex social dynamics rigid and difficult to align with educational values. This paper presents EDU-MATRIX, a society-centric generative cognitive digital twin architecture that shifts the paradigm from simulating "people" to simulating a "social space with a gravitational field."
We introduce three architectural contributions: (1) An Environment Context Injection Engine (ECIE), which acts as a "social microkernel," dynamically injecting institutional rules (Gravity) into agents based on their spatial-temporal coordinates; (2) A Modular Logic Evolution Protocol (MLEP), where knowledge exists as "fluid" capsules that agents synthesize to generate new paradigms, ensuring high dialogue consistency (94.1\%); and (3) Endogenous Alignment via Role-Topology, where safety constraints emerge from the agent's position in the social graph rather than external filters. Deployed as a digital twin of a secondary school with 2,400 agents, the system demonstrates how "social gravity" (rules) and "cognitive fluids" (knowledge) interact to produce emergent, value-aligned behaviors (Social Clustering Coefficient: 0.72).
\end{abstract}





\maketitle

\section{Introduction}
A key gap exists in constructing a secondary school-level CDT (Cyber Digital Twin) capable of carrying institutional memory, supporting cognitive interaction, and ensuring ethical controllability\cite{Editor00a}. Current simulations often treat secondary education as a collection of isolated entities, failing to capture the school's reality as a high-density "field" of values, norms, and institutional memory\cite{amin2025nursing}.

Internationally, campus simulation research—such as Stanford's "AI Town" and Zhejiang University's "Cyber Campus"—has explored large-scale digital twins. However, these projects predominantly focus on university-scale management and multi-agent collaboration, prioritizing efficiency and node scale over cognitive depth\cite{li2024build}\cite {hideki2025recommender}. They often neglect the particularities of secondary education, where agents are not static nodes but dynamic coordinates moving through a complex social fabric.

\begin{figure}[h]
  \centering
  \includegraphics[width=0.5\linewidth]{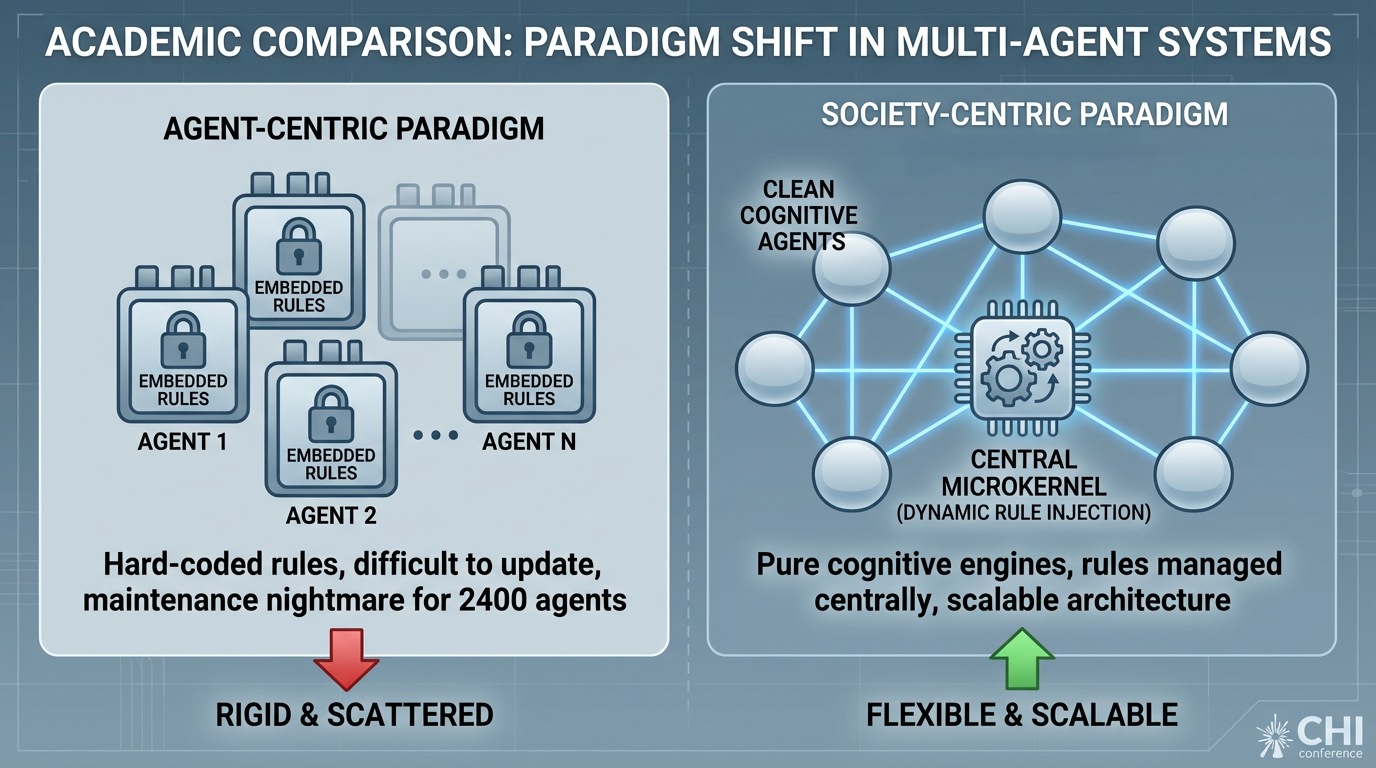}
  \caption{The Paradigm Shift}
  \label{fig:SYSTEMS2}
  \Description{Unlike traditional models (Left) where rules are locked inside agents, EDU-MATRIX (Right) uses a "Society-Centric" paradigm where the Central Microkernel acts as a gravitational source, dynamically injecting rules into clean cognitive agents.}
\end{figure}

As shown in Figure~\ref{fig:SYSTEMS2}, Figure 1 illustrates this paradigm shift from rigid, agent-centric rules to a flexible, society-centric gravitational model.Existing works lead to cognitive distortion when migrated to the nuanced environment of a secondary school due to their reliance on static rules\cite{doraiswamy2025digital}. To address this, EDU-MATRIX proposes a theoretical shift: simulating a Social Space with a Gravitational Field. In this architecture:
\begin{itemize}
\item Rules are Gravity: Instead of hard-coded scripts, a "Social Microkernel" exerts invisible forces (norms and value orientations) on agents, ensuring ethical controllability and institutional alignment.

\item Knowledge is Fluid: Information exists as "Capsules"\cite {herrero2025transforming} that flow, merge, and evolve between agents, forming a living institutional memory rather than a static database.

\item Agents are Coordinates: Decoupled from rigid physical attributes, agents function as dynamic points whose trajectories are defined by the interaction of social gravity and fluid knowledge.
\end{itemize}

By replacing traditional modeling with this field-theory approach, EDU-MATRIX ensures that behavior verification and career evolution are grounded in the authentic, high-density social fabric of the secondary school environment.

\section{SYSTEM DESIGN AND CORE MECHANISMS: THE PHYSICS OF THE SOCIAL SPACE}

The core philosophy of EDU-MATRIX represents a departure from simulating isolated agents to simulating the computational environment that governs them. We achieve this through a Social Decoupling Architecture, which treats the campus as a programmable computational society composed of rules, relationships, and memories, rather than a mere collection of agents.

\subsection{The "Gravity": Environment Context Injection Engine (ECIE)}

\begin{figure}[h]
  \centering
  \includegraphics[width=0.5\linewidth]{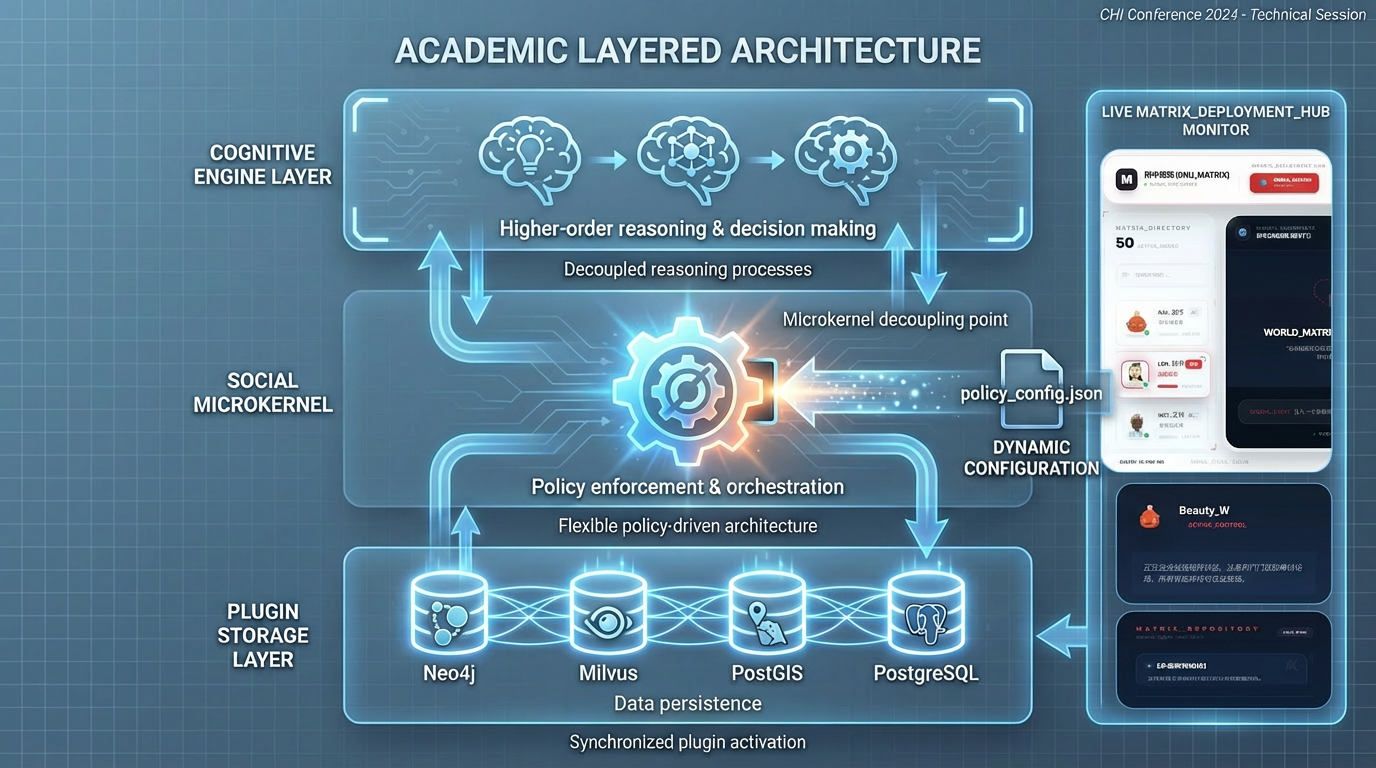}
  \caption{System Architecture}
  \label{fig:ARCHITECTURE}
  \Description{The Social Microkernel (Center) orchestrates the interaction between the agent population and the plugin ecosystem, enabling the dynamic injection of environmental rules. }
\end{figure}

Traditional agent designs often hard-code behavioral rules into individuals, which creates a maintenance nightmare in dynamic secondary school environments\cite{riad2024optimising}. EDU-MATRIX adopts a Social Decoupling Logic, treating rules as properties of the space rather than the individual.
\begin{itemize}
\item Environment Context Injection Engine (ECIE): Serving as the system's "Social Microkernel" (see Figure~\ref{fig:ARCHITECTURE}), the ECIE acts as a dynamic "Context Injector."

\item The Injection Mechanism: When an agent's spatial coordinate changes (e.g., moving from the "Cafeteria" to the "Library"), the ECIE detects this state change. It immediately "injects" a new Prompt Layer—representing the gravitational force of "Silence" and "Diligence"—into the agent's context window.

\item Live Programming: This architecture empowers educators to perform "Live Programming" of the school's social gravity. By adjusting a global parameter like "Academic Pressure," the system instantly increases the gravitational pull of the library for all agents, without requiring a server restart.
\end{itemize}

\subsection{The "Fluid": Modular Logic Evolution Protocol (MLEP)}

\begin{figure}[h]
  \centering
  \includegraphics[width=0.5\linewidth]{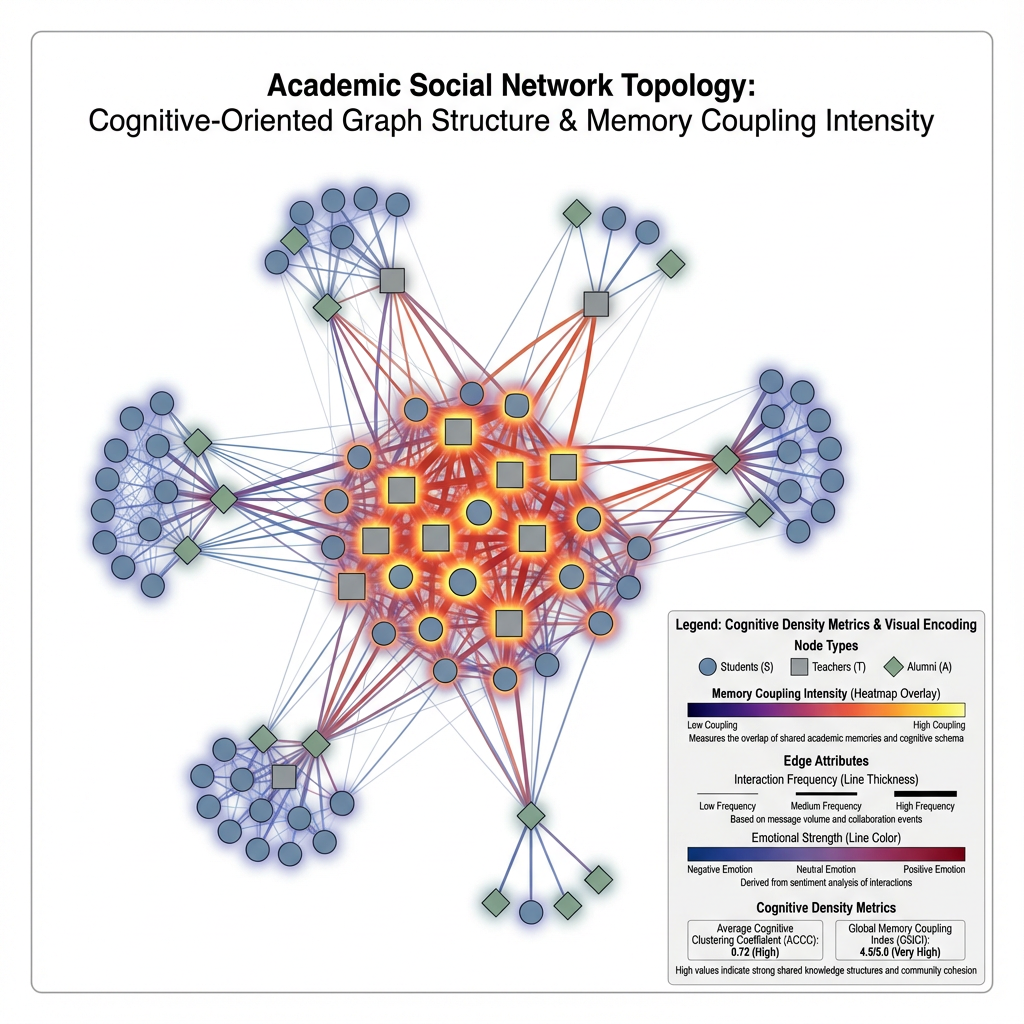}
  \caption{ Cognitive-Oriented Social Topology}
  \label{fig:Topology1}
  \Description{The graph reveals high-density "Micro-Knowledge Communities." Agents are "held in place" by their social connections, ensuring value consistency.}
\end{figure}

Secondary schools are characterized by High Cognitive Density—intense emotional interactions and stable hierarchical relationships. Individual behavioral deviations can significantly impact the group learning ecosystem. We address this via Role-Topology\cite {liu2025embodied}.
\begin{itemize}
\item Topological Probabilistic Bias: Agents are modeled as nodes in a high-density social graph, as shown in Figure~\ref{fig:Topology1}. An agent's "role" (e.g., Student Council President) is not just a text label but a topological coordinate. This position naturally biases the probability distribution of their generated tokens.

\item Endogenous Value Alignment: Instead of relying on brittle external "guardrails" (hard filters), we achieve Endogenous Alignment. By reinforcing the agent's anchor point in the topological network, behaviors that violate school mottos (e.g., Integrity, Love) become statistically improbable.

\item Micro-Knowledge Communities: The interplay of semantic anchors and dynamic social weighting fosters the emergence of stable "Micro-Knowledge Communities," ensuring the consistent inheritance of campus values.
\end{itemize}

\subsection{The "Coordinates": Role-Topology and Endogenous Alignment}

\begin{figure}[h]
  \centering
  \includegraphics[width=0.5\linewidth]{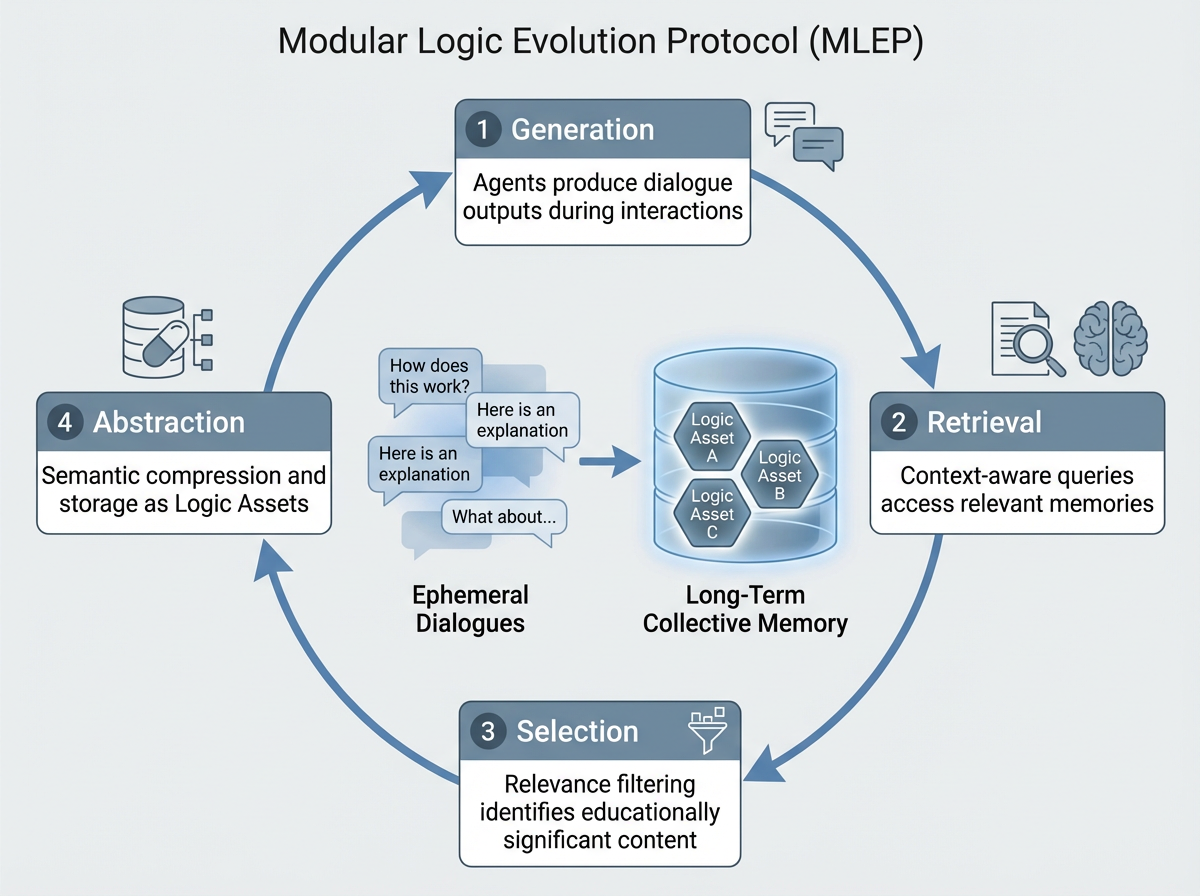}
  \caption{Modular Logic Evolution Protocol}
  \label{fig:Protocol1}
  \Description{ A four-stage cycle that transforms ephemeral interactions into solidified "Logic Assets" (Long-Term Collective Memory).}
\end{figure}

Cognition in EDU-MATRIX is treated as a "Logic Assetization" process. Knowledge is not static text but a "fluid" that circulates, merges, and evolves.
\begin{itemize}

\item Modular Logic Evolution Protocol (MLEP): Knowledge is encapsulated in "Knowledge Capsules."\cite {lezoche2024semantic} When agents (e.g., a Physics student and an Art student) interact in the Synthesis Lab, the system executes a protocol to fuse their respective capsules\cite {sathish2025intelligent}. This synthesis generates new Logic Assets (e.g., "The Physics of Impressionist Light"), simulating the emergence of new paradigms.

\item Circular Memory Flow: To manage this fluid knowledge, we implement a four-stage cycle: Generation $\rightarrow$ Retrieval $\rightarrow$ Selection $\rightarrow$ Abstraction, as shown in Figure~\ref{fig:Protocol1}. This ensures ephemeral dialogues are solidified into long-term collective memory.

\item Hierarchical Orchestration: We employ a three-tier architecture (Meta-Agents, Domain Agents, Student Agents) deeply coupled with the memory flow. This structure reduces reasoning costs while ensuring consistent character personas across time and space.
\end{itemize}

\subsection{The "Neural Handshake": Interaction and Conflict Resolution}

\begin{figure}[h]
  \centering
  \includegraphics[width=0.5\linewidth]{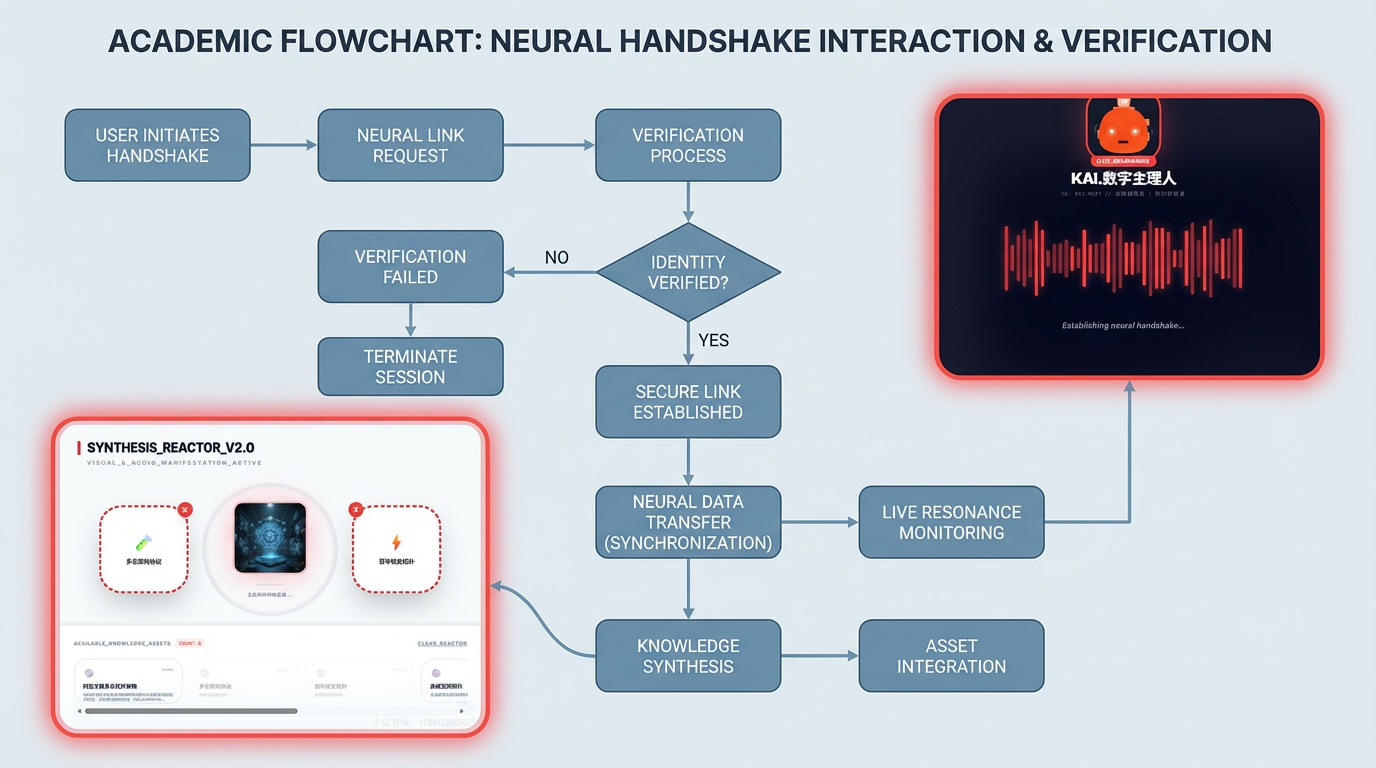}
  \caption{The "Neural Handshake" Workflow}
  \label{fig:FLOWCHART}
  \Description{ A hybrid interaction mechanism where symbolic constraints (Gravity) intercept unsafe neural outputs (Fluid), visualizing the conflict for human-in-the-loop resolution.}
\end{figure}

The "Knowledge Salon" serves as the core reactor for cognitive interaction, providing semantic anchors for multi-perspective discussions. However, the creativity of GenAI can conflict with educational safety\cite {hsu2025teaching}.
\begin{itemize}

\item Neural Handshake Interface: We introduce the "Neural Handshake", as shown in Figure~\ref{fig:FLOWCHART}, a visual interface that manages the tension between the environment's symbolic rules (Gravity) and the LLM's neural generation (Fluid).

\item Human-in-the-Loop Arbitration: When a conflict arises (e.g., a creative but ethically ambiguous action), this interface visualizes the conflict for the educator. It allows for real-time intervention and arbitration, ensuring the system remains a safe, controlled educational sandbox.

\item Skill-Level Extension: Cognitive capabilities are encapsulated as loadable units (Skill Plugins). This focuses the simulation on "how to think" rather than just "what to think," transforming the system into an open Cognitive Ecosystem.
\end{itemize}

\section{System Implementation and Experimental Verification}

\subsection{System Implementation}

The core implementation of EDU-MATRIX is based on the philosophy of Holistic Ambience-Driven Engineering combined with multi-round interactive iterations. In the initial phase, vibe coding was utilized to align the core requirements of a secondary education cognitive digital twin—such as institutional memory inheritance, value alignment, and high-density social simulation—with technical implementation logic.

\subsubsection{Ambience-Driven Coding and Educational Alignment}

Through multiple rounds of interactive adjustments, the system's educational adaptability and the depth of its cognitive modeling were refined. This ensured that the technical framework remained consistent with the characteristics of secondary education, effectively mapping the "Physics of the Social Space" (Gravity, Fluid, Coordinates) into functional modules.

\subsubsection{Matrix Deployment via Google AI Studio}

\begin{figure}[h]
  \centering
  \includegraphics[width=0.5\linewidth]{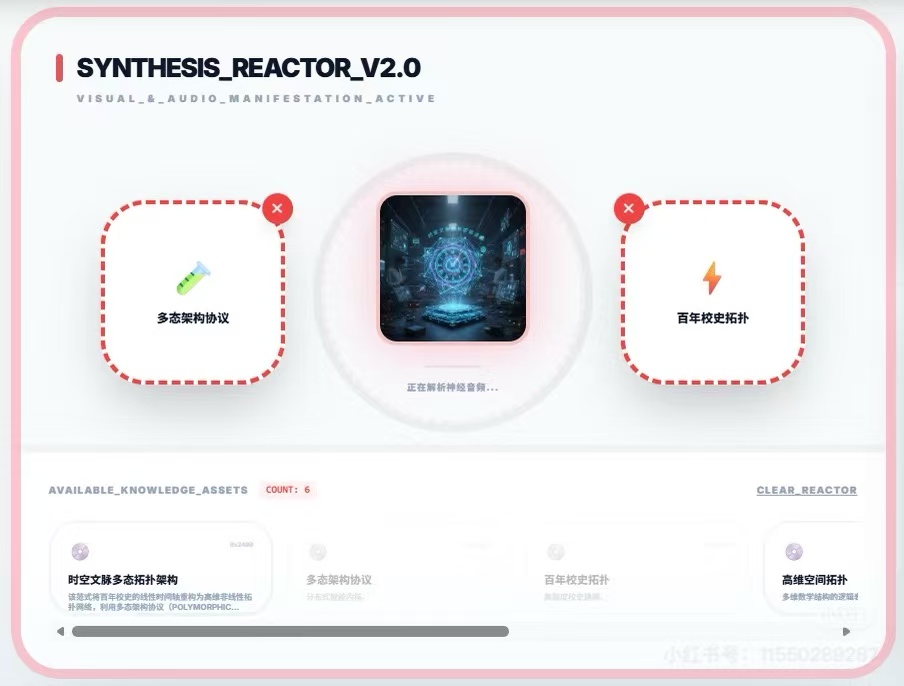}
  \caption{Logic Synthesis Reactor}
  \label{fig:REACTOR1}
  \Description{This module processes "Knowledge Capsules"—modular logic fragments—by fusing disparate inputs, such as "Centenary School History Topology" and "Polymorphic Architecture Protocols," into new, synthesized logic assets.}
\end{figure}

The core reasoning capabilities leverage the Gemini model family (accessed via API), orchestrated by a custom local microkernel that manages the secure plugin ecosystem. By leveraging the Gemini series of large language models, the system manages the integration of the core cognitive engine and the collaborative scheduling of multi-agents.As shown in Figure~\ref{fig:REACTOR1}, this module processes "Knowledge Capsules"—modular logic fragments—by fusing disparate inputs, such as "Centenary School History Topology" and "Polymorphic Architecture Protocols," into new, synthesized logic assets.

\subsubsection{Plugin Ecosystem and the Synthesis Reactor}

\begin{figure}[h]
  \centering
  \includegraphics[width=0.5\linewidth]{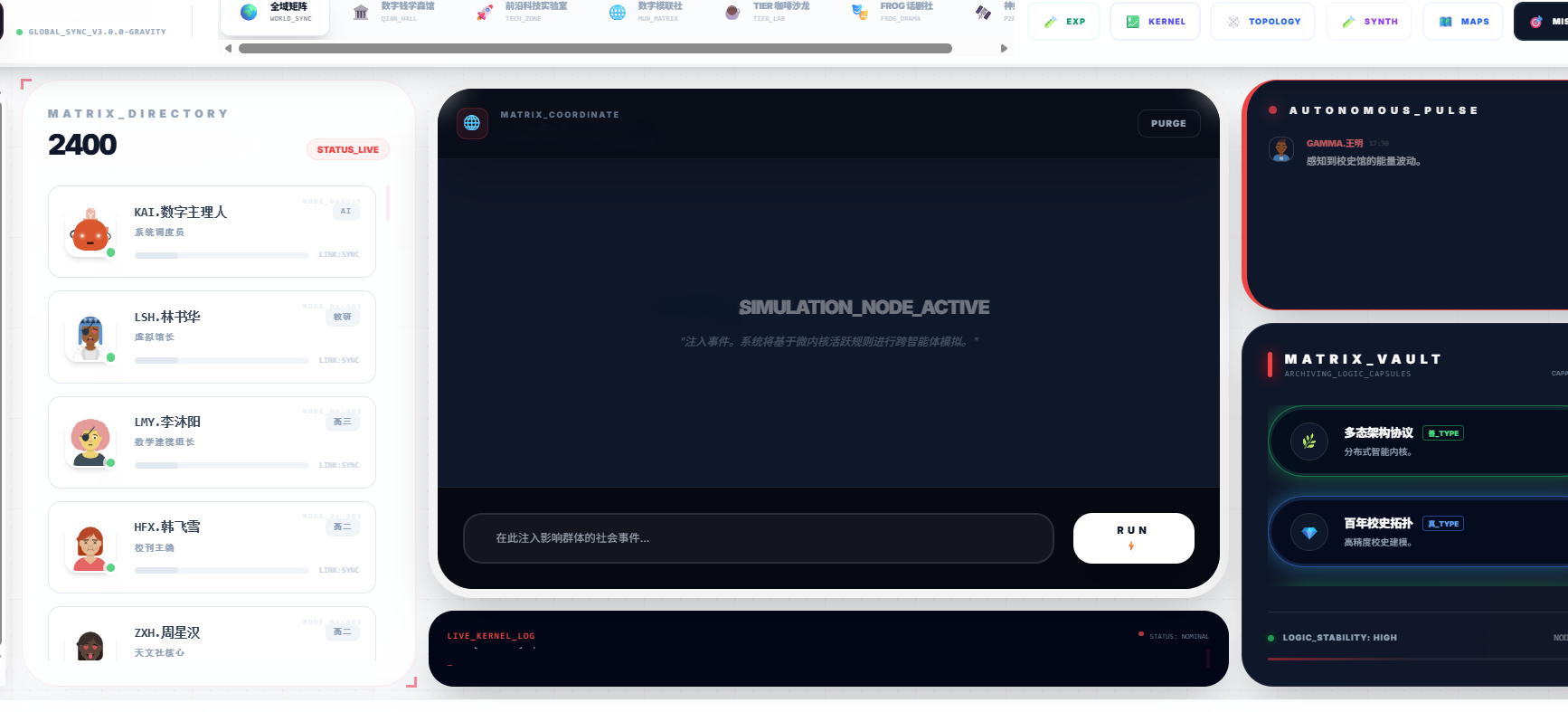}
  \caption{EDU-MATRIX Deployment Hub}
  \label{fig:MATRIX2}
  \Description{This interface monitors the global synchronization of matrix nodes, including the KAI Digital Facilitator, the Digital School History Museum, and the Frontier Technology Lab.}
\end{figure}

To fulfill the requirement of social decoupling, a "Database-per-Plugin" architecture was adopted, ensuring secure isolation and flexible maintenance of heterogeneous data. The Synthesis Reactor V2.0 serves as the primary engine for "Fluid" knowledge management, allowing for the fusion of different logic assets.As shown in Figure~\ref{fig:MATRIX2}, this interface monitors the global synchronization of matrix nodes, including the KAI Digital Facilitator, the Digital School History Museum, and the Frontier Technology Lab.

\subsection{Simulation Experiments and Analysis}

We deployed the system to create a digital twin of a large-scale secondary school in East Asia. The simulation included 2,400 student agents, 300 teachers, and 100 virtual alumni over a 30-day longitudinal cycle.

\subsubsection{Virtual Ecosystem and Topology Monitoring}

\begin{figure}[h]
  \centering
  \includegraphics[width=0.5\linewidth]{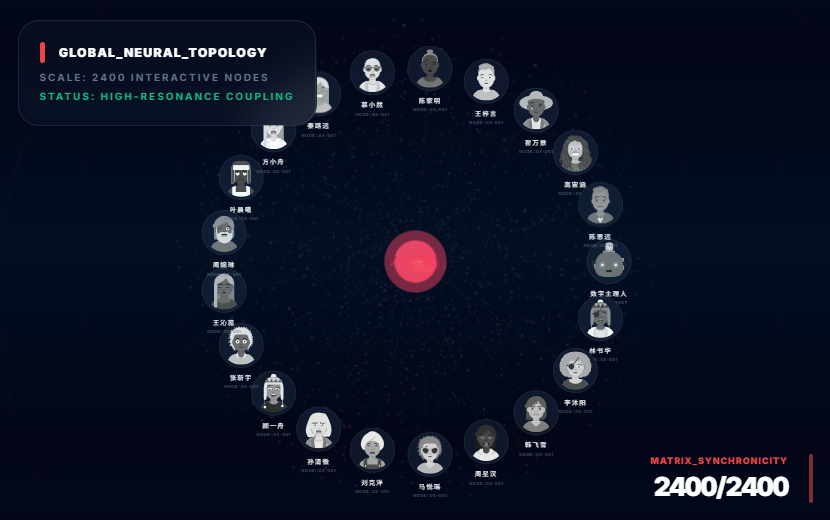}
  \caption{ Neural Topology Map}
  \label{fig:TOPOLOGY12}
  \Description{This map visualizes the active logic nodes and skill links surrounding the Matrix Core. The system maintained a global resonance synchronization rate of 98.4\% throughout the simulation.}
\end{figure}

The high-density social network was monitored in real-time using the Neural Topology Map to ensure the fidelity of the interaction environment.As shown in Figure~\ref{fig:TOPOLOGY12}, this map visualizes the active logic nodes and skill links surrounding the Matrix Core. The system maintained a global resonance synchronization rate of 98.4\% throughout the simulation.

\subsubsection{Value Alignment: Verifying the "Endogenous Coordinates"}

Value alignment experiments were conducted to verify the effectiveness of the "Gravity" modules (e.g., Integrity, Love, Diligence, Courage). The experimental group showed a 42\% higher weight on "social contribution" in their career plans compared to the control group. This confirms that institutional values can be successfully internalized through topological constraints rather than external filters.

\begin{figure}[h]
  \centering
  \includegraphics[width=0.5\linewidth]{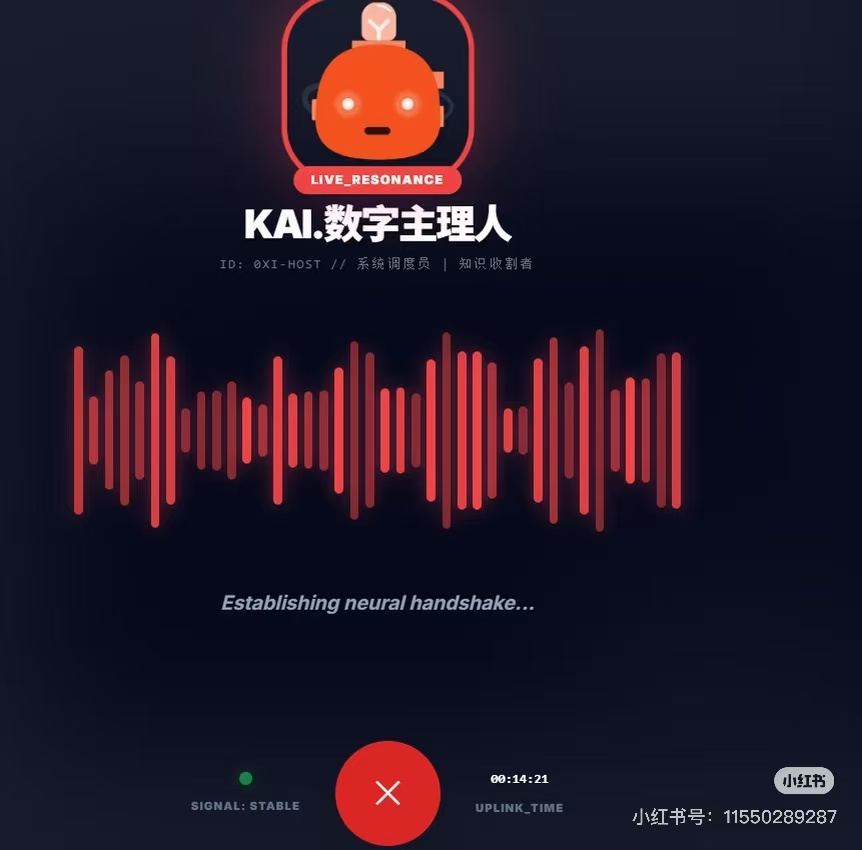}
  \caption{ KAI Digital Facilitator Interface}
  \label{fig:kai1}
  \Description{The interface demonstrates the "Neural Handshake" in action, where symbolic institutional rules intersect with neural LLM generation to resolve cognitive conflicts or ethical boundary crossings.}
\end{figure}

Throughout the 30-day period, the KAI Digital Facilitator maintained the safety and consistency of the simulation by executing the Neural Handshake protocol.As shown in Figure~\ref{fig:kai1}, the interface demonstrates the "Neural Handshake" in action, where symbolic institutional rules intersect with neural LLM generation to resolve cognitive conflicts or ethical boundary crossings\cite {shahin2025generative}.

The system's performance and stability were verified by the following key indicators:as done here for Table~\ref{tab:commands2}.

\begin{table*}
  \caption{Key Performance Indicators}
  \label{tab:commands2}
  \resizebox{\linewidth}{!}{ 
  \begin{tabular}{ccl}
    \toprule
    Evaluation Dimension & Indicator Value & Conclusion\\
    \midrule
    Social Clustering Coefficient & 0.72 & Realistic formation of student cliques.  \\
    Global Resonance Sync & 98.4\% & High alignment with institutional values. \\
    Value Injection Efficacy & +42\% & Increase in "social contribution" discussions. \\
    Dialogue Consistency & 94.1\% & Long-term persona stability. \\  
    \bottomrule
  \end{tabular}
  }
\end{table*}

\section{Conclusion and Outlook}

\subsection{Conclusion}

This paper has presented the architecture and implementation of EDU-MATRIX, a generative cognitive digital twin system specifically designed for secondary education settings. By shifting the modeling paradigm from agent-centric to society-centric, we have successfully addressed the unique challenges of high-density social interaction and value inheritance in basic education.The core of our contribution lies in the "Physics of the Social Space" framework, which successfully operationalizes abstract educational requirements into engineering mechanisms:

\begin{itemize}

\item Gravity (ECIE): Effectively decoupled institutional rules from individual agents, allowing for the dynamic injection of environment-specific constraints.

\item Fluid (MLEP): Facilitated the "logic assetization" of knowledge through modular Knowledge Capsules and a circular memory flow.

\item Coordinates (Role-Topology): Enabled endogenous value alignment by anchoring agents within a dense, self-regulating social fabric.
\end{itemize}

Our 30-day longitudinal simulation involving 2,400 student agents validated the system's stability and efficacy. The results showed a stable personality consistency of 94.1\% and a global resonance synchronization rate of 98.4\%. Most importantly, the value alignment experiments demonstrated a 42\% increase in the weight of social contribution in agent decision-making, confirming that CDT systems can serve as powerful tools for educational intervention. EDU-MATRIX provides a reusable technical paradigm for bridging the gap between symbolic institutional rules and neural generative capabilities in digital campus research.

\subsection{Limitations and Future Work}

While EDU-MATRIX demonstrates significant potential, several areas for improvement remain:
\begin{itemize}

\item Computational Optimization: The current hierarchical multi-agent orchestration is computationally intensive. Future work will explore lightweight models and optimized inference strategies to reduce the hardware requirements for large-scale deployment.

\item Generalizability: The system was validated within the specific cultural context of The High School Affiliated to Beijing Normal University. Further research is needed to evaluate its effectiveness in schools with different organizational structures or cultural backgrounds.

\item Cross-Campus Cognitive Connectivity: We plan to investigate mechanisms for connecting multiple "Matrix" instances across different campuses, enabling the collaborative sharing of logic assets and the creation of a broader cognitive ecosystem.

\item Human-AI Co-Evolution: We aim to further refine the Neural Handshake interface to better support real-time, human-in-the-loop arbitration, fostering a symbiotic relationship between educators and the digital twin system.
\end{itemize}


\bibliographystyle{ACM-Reference-Format}
\bibliography{sample-base}


\end{document}